# Measuring the Electronic Bandgap of Carbon Nanotube Networks in Non-ideal *p-n* Diodes

*Gideon Oyibo[1], Thomas Barrett[1], Sharadh Jois[1], Jeffrey L. Blackburn[2], Ji Ung Lee[1]\**

[1] College of Nanoscale Science and Engineering, State University of New York Polytechnic Institute, Albany, New York 12203, United States.

[2] National Renewable Energy Laboratory, Golden, Colorado 80401, USA

ABSTRACT: The measurement of the electronic bandgap and exciton binding energy in quasi-one dimensional materials such as carbon nanotubes is challenging due to many body effects and strong electron – electron interactions. Unlike bulk semiconductors, where the electronic bandgap is well known, the optical resonance in low dimensional semiconductors is dominated by excitons, making their electronic bandgap more difficult to measure. In this work, we measure the electronic bandgap of networks of polymer-wrapped semiconducting single-walled carbon nanotubes (s-SWCNTs) using non-ideal *p-n* diodes. We show that our s-SWCNT networks have a short minority carrier lifetime due to the presence of interface trap states, making the diodes non – ideal. We then use the generation and recombination leakage currents from these non – ideal diodes to measure the electronic bandgap and excitonic levels of different polymer-wrapped s-SWCNTs with varying diameters: arc discharge (~1.55nm), (7,5) (0.83nm), and (6,5) (0.76nm). Our values are consistent with theoretical predictions, providing insight into the fundamental properties of networks of s-SWCNTs. The techniques outlined here demonstrate a robust strategy that can be applied to measuring the electronic bandgaps and exciton binding energies of a broad variety of nanoscale and quantum-confined semiconductors, including the most modern nanoscale transistors that rely on nanowire geometries.

The 'electronic bandgap' of a semiconductor is defined as the energy difference between the lowest-energy single-particle electron and hole levels. This fundamental property defines many technologically critical properties and processes of a semiconductor, including exciton binding energies, reduction and oxidation potentials in (photo)catalytic reactions, and achievable ranges for quasi-Fermi level splitting in solar cells. In transistors and diodes, the bandgap can determine the leakage current of these devices, an important metric that can determine the efficiencies of the systems they enable. Measuring the electronic bandgap, also known as the transport bandgap of semiconducting single-walled carbon nanotubes (s-SWCNTs), is challenging. Traditional techniques such as optical absorption do not work because of the weak oscillator strength of band-to-band transitions in quasi-one dimensional materials. Instead, the optical absorption is dominated by intense excitonic transitions that arise from the strong coulomb binding between electrons and holes.[1] The use of scanning tunneling spectroscopy poses other problems due to screening from the metal substrates used in these measurements.[2] Here, we measure the electronic bandgap of polymer-wrapped s-SWNTs by creating p-n diodes with an ideality factor of 2, one of only two types of diodes that permits one to extract the bandgap from transport properties. The technique we show and the conclusions we draw are broadly applicable. For example, we note that while s-SWNTs may represent the extreme limit of one-dimensional confinement, the same challenges in measuring the bandgap in s-SWNTs will also apply to current and future transistors as they are shaped into nanowire geometries to allow continued scaling.[3,4]

The use of polymer wrapping in the purification and sorting of s-SWCNTs has given unprecedented access to highly homogeneous chiralities with varying diameters and bandgaps.[5,6] Polymer-wrapped s-SWCNTs are solution processable and are already being used in the fabrication of carbon nanotube microprocessors,[7,8] solar cells,[9–13] thermoelectrics[14,15] and light emitting devices.[16,17] Despite their potential for widespread use, there remain gaps in our understanding of their fundamental properties. While charge transport in polymer-wrapped s-SWCNT networks is already an established field,[18] their intrinsic transport bandgap has remained elusive despite its fundamental and technological importance. Here, we provide measurements of the electronic bandgap of polymer-wrapped s-SWCNTs.

The *p-n* diode is one of the most fundamental building blocks of optoelectronic and electronic devices and it can also be used to study the bandgaps of semiconductors.[19–21] It is already well known that the optical transitions of carbon nanotubes are dominated by excitons (electron-hole pairs bound by a binding energy, $E_b$).[1] The optical bandgap, which is the first excitonic transition, $E_{11}$ of s-SWCNTs is therefore smaller than the electronic bandgap $E_g$ where $E_g = E_{11} + E_b$. In this work, we fabricate *p-n* diodes using networks of polymer-wrapped s-SWCNTs that allow us to measure the diameter-dependent s-SWCNT electronic bandgaps ($E_g$) and binding energies ($E_b$). Our results are consistent with theoretical and experimental values from previous works on the intrinsic bandgap of s-SWNTs but renormalized by the dielectric environment.[2,22,23]

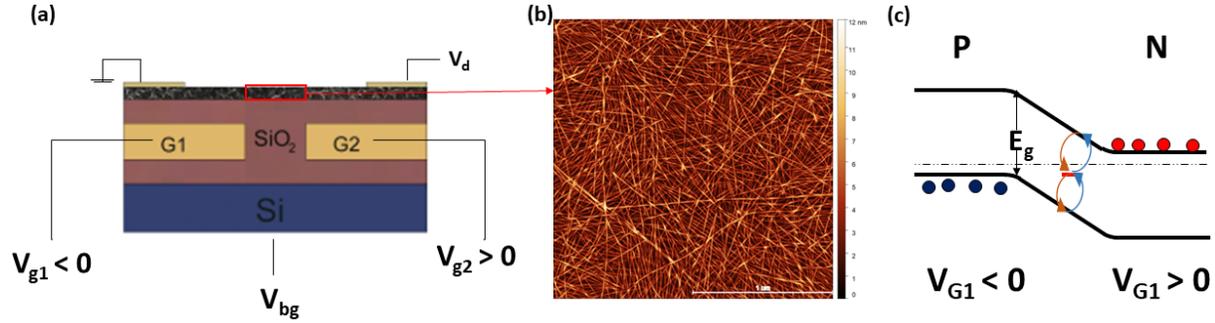

*Figure 1|(a) Device architecture showing p-n diode formed using buried split gates G1 and G2 to electrostatically dope the s-SWCNT network. The back gate $V_{bg}$ is used to ensure ambipolar conduction by measuring the transfer curves (see supplementary Fig. S1). (b) Typical AFM image of s-SWCNT network used in forming p-n diodes. (c) Band diagram of p-n diode showing mid-gap states in the intrinsic region dominating the leakage current ($I_o$) due to generation and recombination of carriers. We use the temperature dependence of $I_o$ to calculate the bandgap of the s-SWNT network.*

**Results and Discussion**

We fabricate *p-n* diodes using electrostatic gating techniques, described in our previous works,[11,24–26] as shown in Fig. 1. Using buried split gates G1 and G2 with split gate spacing, G, ranging from 0.1 $\mu m$ to 0.5 $\mu m$, we apply opposite bias to create *p-* and *n-* doped regions on the s-SWCNT network as shown in the band diagram of Figure 1c. We note that devices with G>0.5 $\mu m$ tend to show diodes with ideality factor *n*>2, with some having *n*>3, which are not suitable for this study. This behavior results from the large disorder present in these films, as we show below.

We characterize *p-n* diodes formed on polymer-wrapped s-SWCNT networks of different diameters. Large-diameter arc discharge s-SWCNTs (~1.55nm) are extracted using PFO-BPy, whereas small-diameter (6,5) (0.76nm) and (7,5) (0.83nm) s-SWCNTs are extracted from CoMoCAT by using PFO-Bpy and PFO respectively.[5,6] Figure 2a shows representative diode current-voltage (*I-V*) curves measured at *T*=300 K from arc, (7,5), and (6,5) networks. We fit the *I-V* characteristics to the diode equation $I_D = I_o \left( \exp\left(\frac{qV}{nKT}\right) - 1 \right)$ to extract their respective leakage currents $I_o$ and ideality factors *n*. *V* is applied voltage and *KT* is thermal energy.

For ideal diodes (*n* = 1), diffusion of minority carriers from the *p*- and *n*- doped regions dominate the reverse bias characteristics, while for non-ideal diodes (*n* = 2), generation and recombination of electron-hole pairs due to mid-gap states in the intrinsic region is responsible for the diode leakage current. We note that these are the only two types of behavior that allow one to measure the bandgap using the thermal activation energy of $I_o$.

In our diodes, the intrinsic region will form between the split-gates. Typically, we observe ideal behavior in nearly abrupt *p-n* diodes with intrinsic spacing of ~0.1 μ*m* and non-ideal behavior with intrinsic spacing greater than ~0.1 μ*m*.

**Generation and Recombination Leakage Current**

We measure the *I-V* characteristics of non-ideal diodes at *T*=300 K as shown in figure 2 with ideality factor, n ~ 2 showing that the reverse bias leakage current in our devices is due to the generation and recombination from mid-gap states in the undoped intrinsic region.

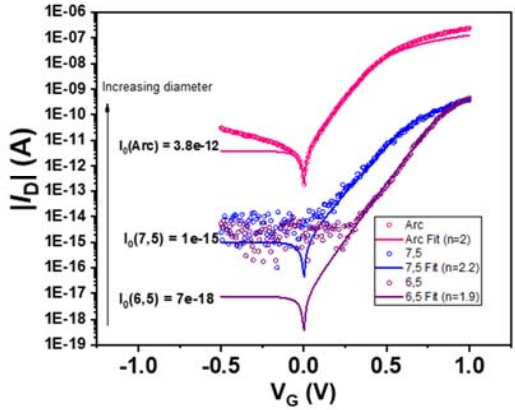 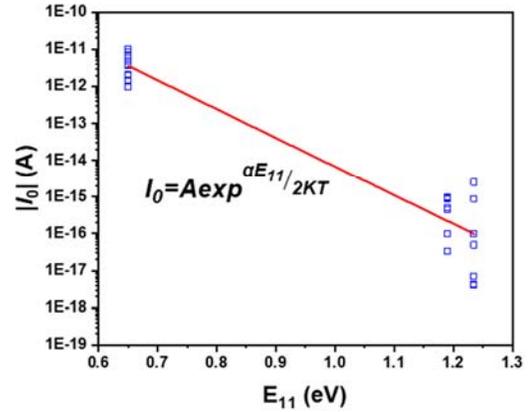

*Figure 2| (a) Dark current voltage (I-V) characteristics of p-n diodes fabricated on networks of different s-SWCNT distributions – arc, (7,5) and (6,5) – showing leakage current ($I_0$) values increasing with diameter (decreasing bandgap) and ideality factor of 2. (b) Leakage current ($I_0$) for all non-ideal diodes fabricated on a natural log scale against the $E_{11}$ optical transitions of the s-SWCNT networks. $E_{11}$ values for our devices are obtained from the photocurrent spectra of the diodes (see Fig. S2)*

We use the phenomelogical Shockley-Read-Hall theory (SRH)[19] to determine the key properties of our polymer-wrapped s-SWCNT films. We assume the films are quasi-two dimensional and use the surface carrier generation rate developed from the SRH model. Using the simplest model where the trap level is at mid-gap, we obtain the surface generation rate $U = n_i/2\tau$. This $U$ gives the highest generation and recombination rate. The intrinsic carrier density is given as $n_i = D_0 e^{-E_g/2KT}$. $D_0$ is related to the effective density of states which we have derived in our previous work for a single nanotube to be $16/3\pi d V_{pp}$, where $d$ is the C-C length and $V_{pp}$ is the hopping energy between the nearest neighbor sites.[27] $\tau$ is the minority carrier lifetime and is inversely related to other parameters, including the trap density and the capture cross section. The leakage current due to generation and recombination, $I_0 = qUW$ can therefore be related to the temperature through the Arrhenius relationship, $I_0 = Ae^{-Eg/2KT}$, where $A$ is a constant, $q$ is the elementary charge, and $W$ is the width of the intrinsic region.[19] We see that the SRH theory naturally gives $Eg/2$ as the activation energy, $Ea$.

We measure the *I-V* characteristics of more than 20 non-ideal diodes across different diameter s-SWCNT networks and plot the leakage current ($I_o$) values against the optical bandgap $E_{11}$ of each s-SWCNT as shown in Fig. 2b. Since both the electronic bandgap, $E_g$, and optical bandgap, $E_{11}$, scale approximately inversely with diameter,[23,26] they are related. We thus expect $I_o$ to depend in a similar way to the optical gap, further supporting that it is a measure of the fundamental properties of s-SWCNTs. To confirm, we use $E_g = \alpha E_{11}$ in $I_0 = Ae^{-Eg/2KT}$,[26] where $\alpha$ is a scaling parameter. In Fig. 2b, we fit to a linear slope and extract $\alpha \approx 0.98$. We note that this value is fortuitously close to the value of 1 and that a large scatter in some of the data, due to process variations in nanotube density across the different s-SWCNT networks that affect the leakage current, makes the correlation only approximate. Nevertheless, the correlation in Fig. 2b clearly links $I_o$ to $E_g$ through $E_{11}$.

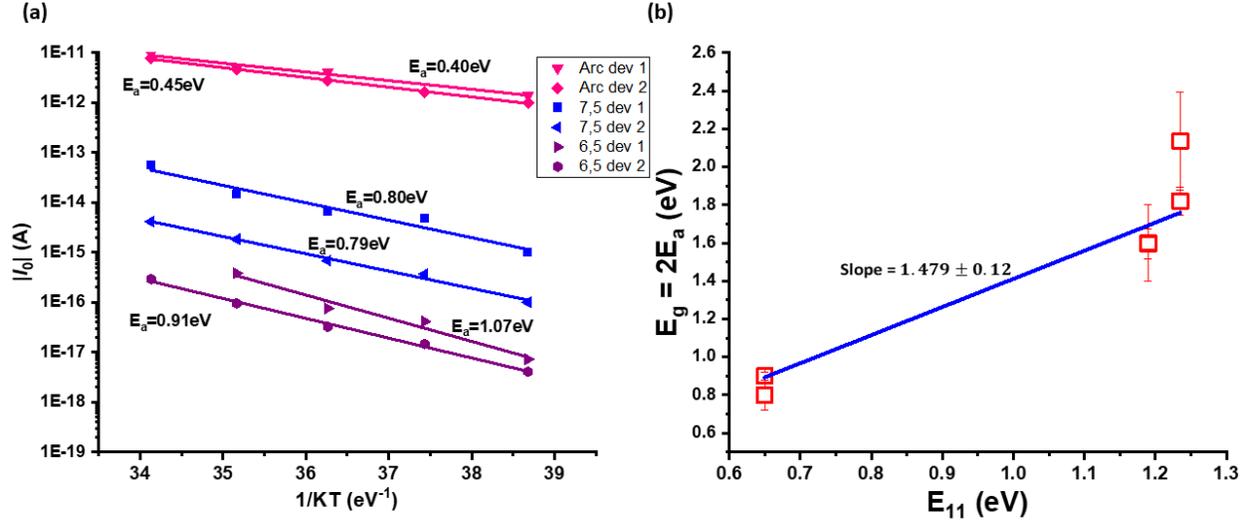

*Figure 3| (a) Leakage current $I_o$ is plotted on a natural logarithmic scale vs 1/KT for two representative devices, each for arc, (7,5), and (6,5) s-SWCNTs. The slope of the linear fit is the activation energy, $E_a$ according to the temperature dependent Arrhenius relationship for $I_0$. (b) Relationship of the measured bandgap $E_g$ ($2E_a$) vs $E_{11}$, a fundamental property of the s-SWCNT network. The error bars account for variations in the scatter plot.*

Temperature – dependent measurements allow us to further clarify the link between the optical and electronic bandgaps. To do so, we use the SRH model for non-ideal diodes ($n$ = 2) to analyze temperature-dependent measurements. Diode *I-V* curves at different temperatures (300K – 340K) are fitted to the dark diode equation with leakage current values extracted as explained above (see Fig. S3). $I_o$ is then plotted on a natural log scale vs 1/KT and the slope fitted to get the activation energy according to the relationship $I_0 \propto e^{\frac{-E_a}{KT}}$ as shown in Fig. 3a. In Fig. 3b, we show that the measured electronic bandgap ($2E_a$) is related to the optical bandgap $E_{11}$. Using $E_{11}$ values from the photocurrent spectra of our devices (see Fig. S2), we derive $E_g \sim 1.48 E_{11}$, which is close to the results from our single-nanotube p-n diode studies[26,27] and provides a relationship between the electronic and optical bandgaps.

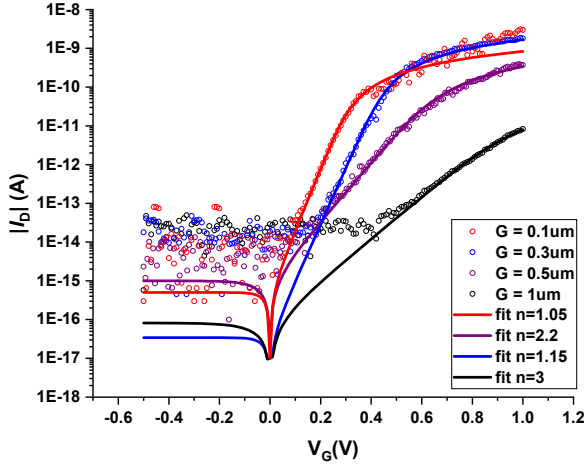

*Figure 4| Dependence of the dark I-V characteristics on the width of the intrinsic region (G) of (7,5) s-SWCNT network p-n diodes. We obtain ideal diode behavior when the intrinsic spacing is 0.1um (nearly abrupt). As the width increases, the diodes become non-ideal.*

Before we continue, we explain our choice to exclude devices with G > 0.5 *μm* and support it by extracting parameters from $U$. To extract parameters from $U$, we scale the $n_i$ derived from above to the maximum potential $n_i$ value that assumes a close-packed aligned array of s-SWCNTs fills the device area. This scaling helps us to approximate the number of nanotubes that contribute to the generation current. Equating $U$ to the generation rate from the leakage currents, we arrive at $\tau \sim 10^{-10} s$. In our calculation, we assume a region about 0.1μm between the gates contributes to the generation of minority carriers. To support this approximation, we show in Fig. 4 the I-V characteristics of (7,5) network devices as a function of the split gate spacing. Some variation in leakage current is expected due to the variation in network thickness across devices. The most important change is the abrupt increase in the ideality factor for split gate spacing greater than 0.5 µm. The minority carrier lifetime $\tau$ that we calculate is very short compared to some of the most pristine interfaces in semiconductor devices. For example, the SiO$_2$/Si interface of a MOSFET is known to have minority carrier lifetimes on the order of 0.1-1 ms.[28] The short lifetime of s-SWNTs implies a large interface trap density and/or capture cross section and is consistent with the rapid transition to n>2 diodes as the width of the intrinsic region increases, as shown in Fig. 4. Since our analysis requires using diodes with $n\sim 2$, we chose diodes with relatively small intrinsic lengths.

Previously, we reported on large interface states that arise from the substrate on sparse network of s-SWNTs.[29] The surface contributes an effective volume around the nanotube that can contribute to trap states. Assuming the surface states are independent of nanotube density, the substrate induced trap states per nanotube decrease with increasing nanotube coverage. As such, we expect the substrate effects to be minimized here, but not absent, since there is a significant coverage from the network. In addition, polymer wrapping and excess polymer may contribute to additional trap states.

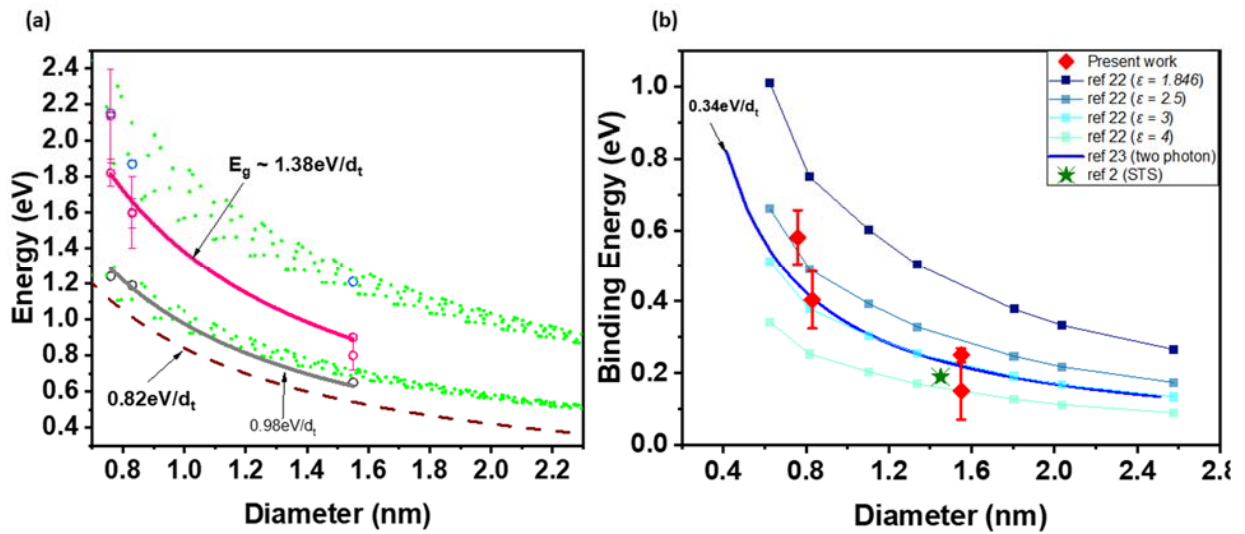

*Figure 5| (a) an updated Kataura plot showing theoretical exciton transitions (closed green circles), measured $E_{11}$ (open gray circles) and $E_{22}$ (open blue circles) transitions from our s-SWCNT network p-n diodes (see Fig. S1), the tight binding bandgap (0.82eV/ $d_t$) and measured bandgap values of our s-SWCNT network (open pink circles). We find a best fit of $E_g \sim 1.38eV/d_t$. (b) Binding energy dependence on diameter also showing data from other published works.*

Next, we examine the extracted bandgap in the broader context of s-SWCNT studies. We show in Fig. 5a the Kataura plot with updated values for the $E_{11}$, $E_{22}$ and $E_g$ of our s-SWCNT networks. We fit our $E_g$ and $E_{11}$ values and derive the relation $E_g \sim 1.38$eV/ $d_t$ and $E_{11} \sim 0.98$eV/ $d_t$, respectively, where $d_t$ is the nanotube diameter. We also show the binding energies calculated for our s-SWCNT networks and compare with previous theoretical[22] and experimental[2,23] works in Fig. 5b. According to Capaz et al, the binding energy of the $E_{11}$ excitons is related to the diameter, $d_t$ through the function, $E_b = \frac{1}{d_t}(A + \frac{B}{d_t} + C\xi + D\xi^2)$, where $\xi = (-1)^v \cdot \cos(\frac{3\theta}{d_t})$. Since we do not observe the chirality dependent effect, $v = (n - m) \ mod \ 3$ in our experimental results, we use the relation $\xi = \cos(\frac{3\theta}{d_t})$ for our comparison. $A = 0.6724$ eV nm, $B = -4.910 \times 10^{-2}$ eV nm$^2$, $C = 4.577 \times 10^{-2}$ eV nm$^2$ and $D = -8.325 \times 10^{-3}$ eV nm$^3$ for nanotubes with dielectric environment, $\varepsilon = 1.846$. For the binding energy dependence on dielectric environment, we adopt the scaling proposed by Perebeinos et al $E_b \propto \varepsilon^{-1.4}$.[30] We observe that the binding energy values of our network is best approximated with a dielectric environment, $\varepsilon$ ranging from 2.5 – 4. This range is consistent with the two-photon excitation spectroscopy measurements of Dukovic et. al. for s-SWCNT thin films embedded in a polymer matrix, where they derived the relation, $E_b = 0.34$eV/$d_t$, consistent with binding energy values in a dielectric environment of $\varepsilon = 3$ (ref. 22). Also, Lin et al used scanning tunneling spectroscopy (STS) to measure the binding energy of a semiconducting single nanotube at a height of ~3.5nm from their metallic substrate, separated by bundles of arc discharge nanotubes (ref. 2).[2]

Finally, we contrast our results to the single nanotube p-n diode studies where we observe ideal diode behavior and bandgap values that are below the $E_{11}$ values. The small bandgap values measured in those studies arise from a significant doping-induced bandgap renormalization in the doped regions.[26,31] In this work, we have measured the bandgap of largely undoped s-SWCNT networks, located between the gated regions. This key difference is possible because Coulombic screening lowers the doping per nanotube in the network compared to single nanotube devices, which reduces the amount of bandgap renormalization.[32,33] Therefore, with a larger bandgap in the doped regions of the network devices compared to the single nanotube devices, we are able to measure the electronic properties of the intrinsic region. Also, by focusing on diodes with $n\sim2$, we are guaranteed to measure the intrinsic bandgap, rather than the renormalized bandgap, however small the renormalization may be.

In conclusion, we have measured the electronic bandgap of polymer wrapped s-SWCNT networks by fabricating non-ideal diodes with diode ideality factor $n\sim2$. We show that the activation energy and excitonic levels follow a universal diameter dependence that allows the extraction of the bandgap and exciton binding energies. We analyze our data in the context of bandgap renormalization due to dielectric screening. The techniques we have demonstrated can also be applied to further the understanding of the electronic bandgaps and exciton binding energies of other nanoscale semiconductors like two-dimensional transition metal dichalcogenides.[34–36]

**METHODS**

**Preparation of Semiconducting SWCNT solutions:**

As discussed in more detail in our previous work,[11] large diameter semiconducting SWCNTs were sorted from raw arc discharge SWCNTs (Carbon Solutions Inc.). PFO-BPy (1mg ml$^{-1}$) obtained from American Dye Source was dissolved in 10ml of toluene and mixed with the raw SWCNT in a 1:2 ratio. The solution is then sonicated using a horn tip sonicator in a cool water bath for 30 mins with 1s pulses (Branson digital sonifier) at 70% amplitude. Next, the sonicated solution is centrifuged at 15000rpm for 10 mins (Hermle Z 36 HK centrifuge 221.22 V20 rotor) and the semiconducting supernatant collected and used as is.

The small diameter semiconducting SWCNTs were extracted from CoMoCAT SG65i material (CHASM) using PFO-BPy and PFO (purchased from American Dye Source) for (6,5) and (7,5) SWCNTs respectively. 2mg ml$^{-1}$ of PFO-BPy or PFO is dissolved in toluene and used to disperse (6,5) and (7,5) SWCNTs respectively, from 0.5 mg ml$^{-1}$ of SG65i by tip sonication for 15 min at 40% intensity (Cole-Palmer CPX 750, 1⁄2" tip) in a cool bath of flowing water (~ 18 ºC). Next, the tip sonicated mixtures were immediately centrifuged at 20 °C and 13200 rpm for 5 mins (Beckman Coulter L-100 XP ultracentrifuge, SW-32 Ti rotor) to remove the undispersed soot. The polymer wrapped supernatants (PFO-BPy/(6,5) and PFO/(7,5)) were then centrifuged again at 20 °C and 24100 rpm for 20 hrs to remove excess polymer. The pellet from each of the polymer wrapped (6,5) and (7,5) solutions was then separated from the supernatant and redispersed in toluene. This process (pelleting and redispersion) was repeated until the absorption of the wrapping polymer (either PFO-Bpy or PFO) approached that of the (6,5) or (7,5) S22 excitonic transition after which the final pellet was then re-dispersed in toluene in a heated ultrasonic bath for more than an hour to yield nearly monochiral (6,5)/PFO-Bpy or (7,5)/PFO s-SWCNT solutions.

**Device Fabrication:**

To form a *p-n* diode on networks of SWCNTs, we fabricate the buried split gates in the 300mm wafer fab of the College of Nanoscale Science and Engineering at SUNY Polytechnic Institute in Albany as described previously.[11,25] Briefly, the device is fabricated using standard lithography, deposition and etch techniques in the SUNY Polytechnic Institute 300mm fab. The fabrication starts with a 300mm poly-Si wafer, which is made highly conductive by phosphorous implantation with a concentration of ~ $10^{19}$ cm$^{-3}$ to ~ $10^{20}$ cm$^{-3}$ followed by annealing to activate the dopants, to form the back gate. 100 nm of silicon dioxide ($SiO_2$) dielectric is then deposited over the heavily p-doped silicon wafer using a wet thermal oxidation process. Next, the buried split gates are formed using 100nm of polysilicon deposited over the dielectric and doped by ion implantation. Using standard photolithography and subtractive etch techniques, the polysilicon is etched to define the split gates with inter-gate spacing ranging between 0.1 μm to 1 μm. On top of the split gates, 150nm of $SiO_2$ is further deposited using a plasma enhanced chemical vapor deposition process (PECVD) and polished using chemical mechanical polishing (CMP) to achieve an atomically flat surface until a desired dielectric thickness of 100nm is achieved. The CMP process is precisely controlled by checking the dielectric thickness at multiple intervals. To allow probes to land on the poly-Si bondpads for electrostatic gating, the dielectric above the bondpads is then subsequently etched. Onto these pre-fabricated structures we then deposit the s-SWCNT networks using repetitive immersion and soak in hot toluene at 120C for 10 mins to remove excess polymer.

To complete the device, we use electron beam lithography and oxygen plasma etch to define the s-SWCNT channel and deposit 20nm/20nm of Ni/Au to form the source and drain contacts.

**Measurement Methods:**

All electrical measurements are performed in vacuum ($< 5 \times 10^{-5}$) at temperatures 300K – 340K (for temperature dependent measurements, see figure S3) using an Agilent B1500A semiconductor parameter analyzer. Photocurrent measurements are carried out using an NKT photonics broadband laser dispersed through a monochromator.

**Atomic Force Microscopy (AFM) Measurements:**

AFM measurements for SWCNT thickness were acquired using a Bruker Dimension icon AFM with ScanAsyst and gwyddion 2.6 software used for analysis. Scans were acquired across the SWCNT network with scan rate of 0.2Hz at 1024 × 1024 resolution. Scanasyst-air silicon probe tip was used.

**Supplementary Information.**

The following are available as supporting information:

Transfer curves of arc, (7,5) and (6,5) s-SWCNT network devices; normalized photocurrent spectra of arc, (7,5) and (6,5) s-SWCNT network diodes; fitted temperature dependent current – voltage (I-V) curves for arc, (7,5) and (6,5) diodes.

**Corresponding Author**

*Email: leej5@sunypoly.edu

**Author Contributions**

JUL conceived and designed the study; JB purified the (6,5) and (7,5) CNTs; GO fabricated the devices, performed transport and photocurrent measurements, and carried out the data analysis; TB helped with AFM characterization, JUL is the principal investigator. All authors contributed to the preparation of the manuscript.

**Acknowledgment**

We would like to acknowledge the support of NREL for their help in the purification of (6,5) and (7,5) SWCNTs, and CNSE metrology and AESG for on site support with tools. This work was authored in part by the National Renewable Energy Laboratory (NREL), operated by Alliance for Sustainable Energy, LLC, for the U.S. Department of Energy (DOE) under Contract No. DE-AC36-08GO28308. Support for J.B. was provided by the Solar Photochemistry Program, Division of Chemical Sciences, Geosciences, and Biosciences, Office of Basic Energy Sciences, U.S. Department of Energy (DOE).

**TOC GRAPHIC**

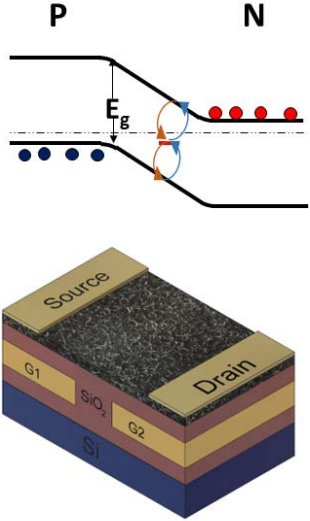